\documentclass{optica-article}

\journal{opticajournal} 

\articletype{Research Article}

\usepackage{lineno}

\begin{document}

\title{Pure moving optical media consisting of magnetochiral metasurfaces}

\author{Toshiyuki Kodama,\authormark{1,*} Toshihiro Nakanishi,\authormark{2} Kei Sawada,\authormark{3,4} and Satoshi Tomita\authormark{1,5,$\dag$}}

\address{\authormark{1} Institute for Excellence in Higher Education, Tohoku University, Sendai 980-8576, Japan\\
\authormark{2} Division of Electronic Science and Engineering, Graduate School of Engineering, Kyoto University, Kyoto 615-8515, Japan\\
\authormark{3} RIKEN SPring-8 Center, Sayo, Hyogo 679-5148, Japan\\
\authormark{4} Cornell High Energy Synchrotron Source, Cornell University, Ithaca, NY 14853, USA\\
\authormark{5} Department of Physics, Graduate School of Science, Tohoku University, Sendai 980-8578, Japan}

\email{\authormark{*}tkodama@tohoku.ac.jp} 
\email{\authormark{$\dag$}tomita@tohoku.ac.jp}


\begin{abstract*} 
We report numerical studies of 
microwave bianisotropies by  
magnetochiral (MCh) metasurfaces 
consisting of double Z-type gammadions 
with perpendicularly magnetized substrates. 
The metasurfaces' effective polarizability tensor, 
extracted from calculated reflection and transmission coefficients, 
has components of 
the non-reciprocal moving-type bianisotropy 
as well as the reciprocal chiral-type bianisotropy 
and non-reciprocal magneto-optical (MO) effect. 
Combination of the metasurfaces with 
contraposition MCh metasurfaces 
cancels both the chiral-type bianisotropy and MO effect, 
resulting in pure moving optical media.
Achieved perfect transmission with phase difference of $\pi$
realizes an ideal gyrator for arbitrary spatial and polarization modes.
\end{abstract*}

\section{Introduction}
Symmetries relate to laws in optics.
Noether's theorem says that 
a system with a continuous symmetry 
has a conserved quantity; 
for example, 
the frequency (energy) of light is conserved 
in a medium with the time-translational symmetry.
Another kind of symmetries is a discrete one, such as 
the spatial-inversion and time-reversal symmetry. 
These symmetries 
impose restrictions on optical phenomena 
and play essential roles in optics \cite{LL}. 
A breaking of the space-inversion and/or time-reversal symmetry
gives rise to bianisotropies \cite{Kong,simovski_book}, 
which represent cross-correlated electromagnetic inductions 
among electric polarizability, magnetization, and electromagnetic fields  
i.e., magnetoelectric couplings, 
in constitutive equations. 
The bianisotropies are classified into the four categories 
associated with reciprocity and polarization dependence as follows \cite{asadchy2018}:
a reciprocal chiral-type bianisotropy with polarization plane rotation,
a reciprocal omega-type without polarization rotation, 
a non-reciprocal Tellegen-type with polarization rotation,
and a non-reciprocal moving-type without polarization rotation.
The control of the bianisotroies is a challenge 
in natural materials with broken symmetries 
such as multiferroic materials \cite{toyoda2015,kuzmenko2015,fiebig2016review};
therefore,
man-made structured materials, 
referred to as metamaterials, 
are suitable to explore bianisotropies \cite{chen2005,kida2005,plum2009,li2009,niemi2013}.
In particular, 
metasurfaces with two-dimensional planar structures 
are good playgrounds 
to embody a tailor-made bianisotropic medium \cite{pfeiffer2014prap,pfeiffer2014prl,odit2016,lavigne2021}.

In this work,
we theoretically and numerically study 
magnetochiral (MCh) metasurfaces 
as non-reciprocal moving-type bianisotropic media. 
The MCh metasurfaces 
consist of double Z-type gammadion with the broken space-inversion symmetry \cite{kim2014} 
and perpendicularly magnetized substrate with the broken time-reversal symmetry. 
The optical MCh effect 
\cite{groenewege1962,portigal1971,baranova1977,markelov1977,baranova1979,wagni1982,wagni1984,barron1984,rikken1997,kleindienst1998,rikken1998,vallet2001,train2008,tomita2014,tomita2018,atzori2020,caridad2021} 
is manifested in 
a transmission coefficient difference, 
which is dependent on propagation direction but independent of polarization,
due to the moving-type bianisotropy. 
We calculate 
transmission and reflection coefficient spectra 
at normal incidence of microwaves 
to evaluate effective polarizability tensor \cite{mirmoosa2014,alaee2015,yazdi2016}.
The effective polarizability tensor 
demonstrates the moving-type bianisotropy 
as well as the chiral-type bianisotropy and magneto-optical (MO) effect.
Furthermore, 
a composite metasurface consisting of 
the ordinary and contraposition MCh metasurcfaces 
realizes pure moving optical media 
without optical activity and MO effect.
The pure moving media show  
the perfect transmission with a phase difference of $\pi$
depending on propagation direction, 
bringing about an ideal gyrator \cite{pozar}.

The pure moving optical media
are of great interest 
in terms of development of functional metasurfaces.
A similar story can be found 
in the development of 
the well-known negative refractive index metamaterials 
consisting of split-ring resonators (SRRs). 
A single C-shaped SRR 
shows the omega-type bianisotropy
due to electromagnetic interaction 
in addition to magnetic permeability \cite{smith2006,kriegler2010}.
This brings about electromagnetic responses 
not only to magnetic field but also to electric field.
An inverse-C-shaped SRR 
is thus incorporated in the inner circle of 
the C-shaped SRR, 
resulting in 
vanishment of the omega-type electromagnetic interaction 
but remaining of the permeability component.
This way, 
the double SRRs 
construct pure negative magnetic permeability media, 
bringing about negative refractive index metamaterials.
Similarly, 
the pure moving medium without the optical activity and MO effect 
obtained in this study 
is a key to realize an ideal gyrator for arbitrary spatial and polarization modes. 

This paper is organized in five sections.
Section 2 details theoretical background, 
including effective polarizability tensor, principle of parameter retrieving, and 
classification of bianisotropic metasurfaces.
Section 3 devotes to numerical calculation setup.
Section 4 describes results and discussion 
of calculated microwave transmission and reflection coefficient spectra 
of the uniaxial bianisotropic MCh metasurfaces.  
The extracted effective polarizability tensor 
of the chiral and MCh metasurfaces 
is illustrated. 
We present pure moving optical media
by introducing contraposition MCh metasurfaces 
with the inverse chirality and converse magnetization 
of the ordinary MCh metasurfaces.
The pure moving media 
result in perfect transmission and phase difference of $\pi$.
Section 5 concludes the paper.

\section{Theoretical background}
\subsection{Effective polarizability tensor}
The electromagnetic properties of a metasurface, 
which interacts with normally incident plane waves 
propagating in the $+z$ direction, 
can be fully characterized by 
the following effective polarizability tensor  $\hat{\alpha}_{ij}$
defined as
\begin{equation}
\begin{pmatrix}
P_{x} \\
P_{y} \\
M_{x} \\
M_{y} \\
\end{pmatrix}
=
\begin{pmatrix}
\hat{\alpha}_{xx}^{\rm ee} & \hat{\alpha}_{xy}^{\rm ee} & \hat{\alpha}_{xx}^{\rm em} & \hat{\alpha}_{xy}^{\rm em} \\
\hat{\alpha}_{yx}^{\rm ee} & \hat{\alpha}_{yy}^{\rm ee} & \hat{\alpha}_{yx}^{\rm em} & \hat{\alpha}_{yy}^{\rm em} \\
\hat{\alpha}_{xx}^{\rm me} & \hat{\alpha}_{xy}^{\rm me} & \hat{\alpha}_{xx}^{\rm mm} & \hat{\alpha}_{xy}^{\rm mm} \\
\hat{\alpha}_{yx}^{\rm me} & \hat{\alpha}_{yy}^{\rm me} & \hat{\alpha}_{yx}^{\rm mm} & \hat{\alpha}_{yy}^{\rm mm} \\
\end{pmatrix}
\begin{pmatrix}
E_{x}^{\rm inc} \\
E_{y}^{\rm inc} \\
H_{x}^{\rm inc} \\
H_{y}^{\rm inc} \\
\end{pmatrix},
\label{eq:bapt}
\end{equation}
where the superscript e and m in $\hat{\alpha}_{ij}$ 
are assigned to electric and magnetic stimulus/responses, respectively.
The subscript $i =x,y$ ($j=x, y$) in $\hat{\alpha}$ 
represents the polarization direction of the output (input) waves. 
The $i$ components ($i = x, y$) of 
induced electric moment per a unit area, 
induced magnetic moment per a unit area, 
incident electric field, 
and incident magnetic field, 
are represented respectively by 
$P_i$, $M_i$, $E^{\rm inc}_i$, and $H^{\rm inc}_i$ \cite{simovski_book}.
In this way, 
$\hat{\alpha}^{\rm em}_{xy}$, for example, represents 
an effective polarizability tensor element, 
where a magnetic field in the $y$ direction 
induces an electric dipole moment in the $x$ direction. 
General metasurfaces 
with anisotropic magneto-electric interactions 
are refereed to as bianisotropic metasurfaces.

Here we write the plane waves to be, 
$E, H \propto {\rm e}^{j (- \vec{k} \cdot \vec{r} + \omega t)}$, 
where $j = \sqrt{-1}$ is the imaginary unit. 
In this notation, 
the effective polarizability tensor components 
can be expressed in a non-dimensional manner to be 
\begin{align}
  \alpha_{xx}^{\rm ee} \omega Z_0 &= \frac{j}{2} (t_{xx}^{+} + t_{xx}^{-} + r_{xx}^{+} + r_{xx}^{-} - 2), \label{eq:XXee}\\
  \alpha_{yy}^{\rm mm} \omega Z_0^{-1} &= \frac{j}{2} (t_{xx}^{+} + t_{xx}^{-} - r_{xx}^{+} - r_{xx}^{-} - 2), \label{eq:YYmm}\\
  \alpha_{xy}^{\rm em} \omega &= \frac{j}{2} (t_{xx}^{+} - t_{xx}^{-} + r_{xx}^{+} - r_{xx}^{-}), \label{eq:XYem}\\
  \alpha_{yx}^{\rm me} \omega &= \frac{j}{2} (t_{xx}^{+} - t_{xx}^{-} - r_{xx}^{+} + r_{xx}^{-}), \label{eq:YXme}\\
  \alpha_{yx}^{\rm ee} \omega Z_0 &= \frac{j}{2} (t_{yx}^{+} + t_{yx}^{-} + r_{yx}^{+} + r_{yx}^{-}), \label{eq:YXee}\\
  \alpha_{xy}^{\rm mm} \omega Z_0^{-1} &= -\frac{j}{2} (t_{yx}^{+} + t_{yx}^{-} - r_{yx}^{+} - r_{yx}^{-}), \label{eq:XYmm}\\
  \alpha_{yy}^{\rm em} \omega &= \frac{j}{2} (t_{yx}^{+} - t_{yx}^{-} + r_{yx}^{+} - r_{yx}^{-}), \label{eq:YYem}\\
  \alpha_{xx}^{\rm me} \omega &= -\frac{j}{2} (t_{yx}^{+} - t_{yx}^{-} - r_{yx}^{+} + r_{yx}^{-}), \label{eq:XXme}
  \end{align}
with transmission coefficients $t^\pm_{ij}$
and reflection coefficients $r^\pm_{ij}$,
where the superscript $\pm$ represents the sign of $z$ component of the wavevector \cite{yazdi2016}. 
A wave impedance of a vacuum is represented by $Z_0$.
The above components are derived from the transmission and reflection coefficients 
for $x$-polarized incident waves, i.e. $t^\pm_{i x}, r^\pm_{i x} (i=x, y)$.
The effective polarizabilities given by Eqs.~\eqref{eq:XXee} to \eqref{eq:YXme} 
are related to polarization-maintaining interaction,
while those given by Eqs.~\eqref{eq:YXee} to \eqref{eq:XXme} 
are related to interaction involving polarization conversion from $x$-polarized waves to $y$-polarized waves. 
The remaining components can be calculated from the transmission and reflection coefficients 
for the $y$-polarized incident waves, i.e. $t^\pm_{i y}, r^\pm_{i y} (i=x, y)$.

Suppose a uniaxial metasurface with the $z$-axis rotational symmetry.
The following relations hold: 
$t_{xx}^{\pm} = t_{yy}^{\pm}$, 
$r_{xx}^{\pm} = r_{yy}^{\pm}$, 
$t_{xy}^{\pm} = - t_{yx}^{\pm}$, 
$r_{xy}^{\pm} = - r_{yx}^{\pm}$, 
which yield 
$\hat{\alpha}_{xx}^{\rm ee} = \hat{\alpha}_{yy}^{\rm ee}$,
$\hat{\alpha}_{xx}^{\rm mm} = \hat{\alpha}_{yy}^{\rm mm}$,
$\hat{\alpha}_{xy}^{\rm em} = - \hat{\alpha}_{yx}^{\rm em}$,
$\hat{\alpha}_{yx}^{\rm me} = - \hat{\alpha}_{xy}^{\rm me}$,
$\hat{\alpha}_{yx}^{\rm ee} = - \hat{\alpha}_{xy}^{\rm ee}$,
$\hat{\alpha}_{xy}^{\rm mm} = - \hat{\alpha}_{yx}^{\rm mm}$,
$\hat{\alpha}_{yy}^{\rm em} = \hat{\alpha}_{xx}^{\rm em}$,
$\hat{\alpha}_{xx}^{\rm me} = \hat{\alpha}_{yy}^{\rm me}$.
As a result, 
the uniaxial metasurface 
is fully characterized 
by the components of the effective polarizability tensor
given by Eq.~\eqref{eq:XXee} to Eq.~\eqref{eq:XXme}. 
In this paper, 
we evaluate these effective polarizabilities
using transmission and reflection coefficients 
from numerically calculated $S$ parameters 
of the MCh metasurfaces.

\subsection{Principle of parameter retrieving}
\begin{figure}[bt!]
\centering\includegraphics[width=7cm]{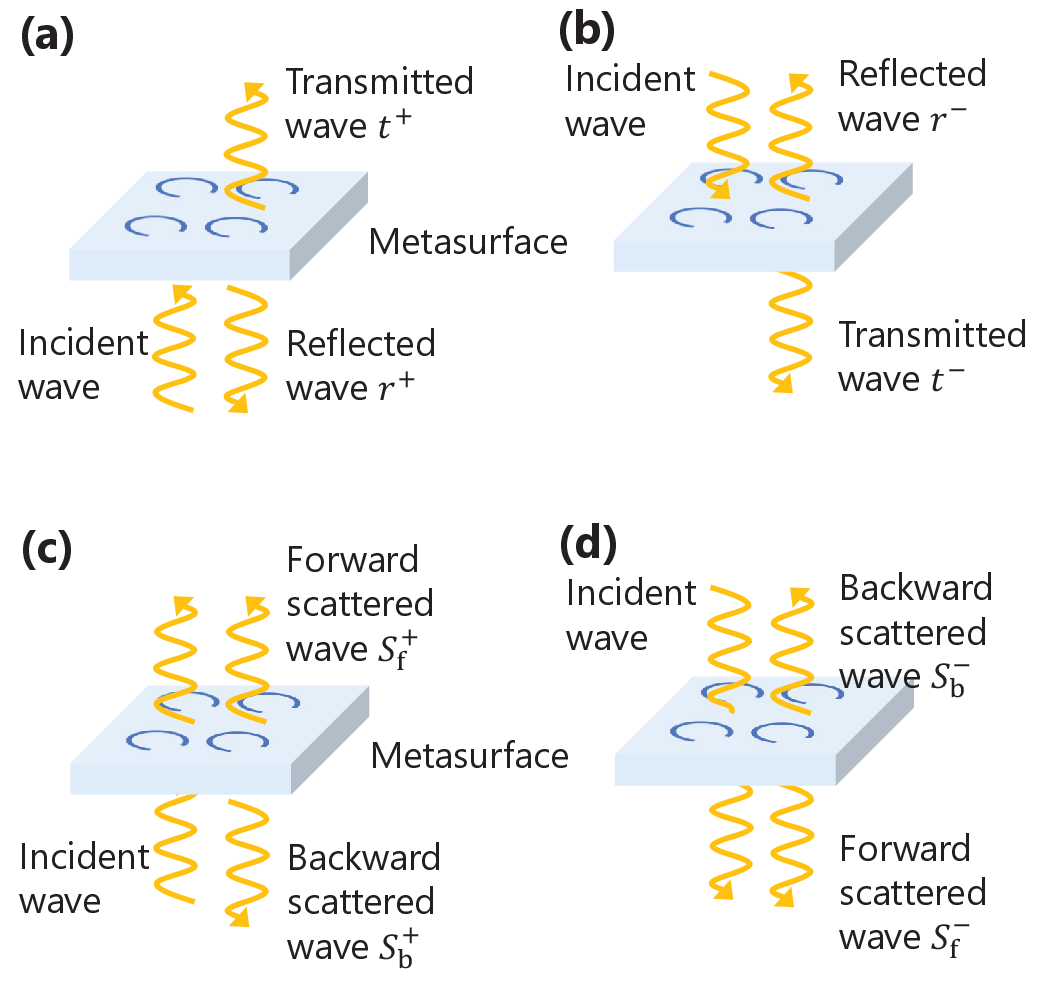}
\caption{
(a) A transmitted wave $t^{+}$ and reflected wave $r^{+}$ 
by the incident wave in the positive direction ($+$).
(b) A transmitted wave $t^{-}$ and reflected wave $r^{-}$ 
by the incident wave in the negative direction ($-$).
(c) Redrawn of (a) using the forward scattered wave $S_{f}^{+}$ 
and the backward scattered wave $S_{b}^{+}$.
(d) Redrawn of (b) using the forward scattered wave $S_{f}^{-}$ 
and the backward scattered wave $S_{b}^{-}$.  }
\label{fig:randt}
\end{figure}

We provide physical insights into 
the parameter retrieving given by Eqs.~\eqref{eq:XXee} to \eqref{eq:XXme}.
For sake of simplicity, 
the case without polarization rotation 
is considered; 
$t^\pm_{xy}=t^\pm_{yx}=r^\pm_{xy}=r^\pm_{yx}=0$.
Moreover, we simplify as 
$t^\pm=t^\pm_{xx}$ and $r^\pm=r^\pm_{xx}$.
Figures \ref{fig:randt}(a) and \ref{fig:randt}(b) 
illustrate electromagnetic responses by 
incident waves propagating in the positive and negative direction, respectively.
Electromagnetic waves in Figs.~\ref{fig:randt}(a) and \ref{fig:randt}(b) 
can be regarded as
superpositions of incident waves and scattered waves  $S^\pm_l (l={\rm f,b})$
radiated by electric and/or magnetic responses of the metasurface
as shown in Figs.~\ref{fig:randt}(c) and \ref{fig:randt}(d).
The superscript, $\pm$, 
represents the direction of the incident waves,
and $S^\pm_{\rm f}$ ($S^\pm_{\rm b}$) 
represents the scattered waves propagating forward (backward)
with respect to the incident waves.
Comparing Figs. \ref{fig:randt}(a) (b) and (c) (d),
we obtain
\begin{align}
t^{\pm} &= 1 + S_{\rm f}^{\pm}, \label{eq:tpm}\\
r^{\pm} &= S_{\rm b}^{\pm}. \label{eq:rpm}
\end{align}

On the metasurface, 
electric and magnetic fields 
coexist for one-way propagating waves.
Nevertheless, 
either the electric fields or magnetic fields 
can be zero 
by destructive interference 
between plane waves with opposite propagation directions,
which form standing waves.
For example, 
when electromagnetic waves with in-phase electric fields propagating in opposite directions
are incident on the metasurface, 
the electric fields are constructively interfered,
and the magnetic fields are completely suppressed by destructive interference. 
As a result, 
only electric fields induce 
electric dipole moment $P$ and magnetic dipole moment $M$
on the metasurfaces, 
resulting in 
$P = \hat{\alpha}^{\rm ee} E^{\rm inc}$ and $M = \hat{\alpha}^{\rm me} E^{\rm inc}$.
Total scattered waves 
can be obtained by adding the scattered waves 
in Figs.~\ref{fig:randt}(c) and \ref{fig:randt}(d):
\begin{align}
  S^{\uparrow} &= S_{\rm f}^{+} + S_{\rm b}^{-} = t^{+} + r^{-} - 1, \label{eq:sup}\\
  S^{\downarrow} &= S_{\rm b}^{+} + S_{\rm f}^{-} = t^{-} + r^{+} - 1, \label{eq:sdown}
\end{align}
where $S^{\uparrow}$ ($S^{\downarrow}$)
represents the scattered waves propagating in upward (downward) direction.
The electric dipole moment $P$ excited on the metasurface 
radiates in-phase electric fields 
whereas the magnetic dipole moment
$M$ radiates out-of-phase electric fields (or in-phase magnetic fields). 
In other words,
a symmetric component of $S^{\uparrow}$ and $S^{\downarrow}$ 
gives electric dipole radiation, 
while an anti-symmetric component of $S^{\uparrow}$ and $S^{\downarrow}$ 
gives magnetic dipole radiation.
Therefore, we obtain  
\begin{align}
\hat{\alpha}^{\rm ee} &\propto S^{\uparrow} + S^{\downarrow} = t^{+} + t^{-} + r^{+} + r^{-} - 2, \label{eq:aee}\\
\hat{\alpha}^{\rm me} &\propto S^{\uparrow} - S^{\downarrow} = t^{+} - t^{-} - r^{+} + r^{-}. \label{eq:ame}
\end{align}
Equation \eqref{eq:aee} corresponds to Eq.~\eqref{eq:XXee}, 
and Eq.~\eqref{eq:ame} corresponds to Eq.~\eqref{eq:YXme}.

Contrastingly, 
magnetic response 
$P = \hat{\alpha}^{\rm em} H^{\rm inc}$ and $M = \hat{\alpha}^{\rm mm} H^{\rm inc}$ 
can be obtained 
by incident electromagnetic waves propagating in opposite directions 
with out-of-phase electric fields.
In this case, 
total scattered waves 
can be derived by subtracting the field in Fig.~\ref{fig:randt}(d)
from that in Fig.~\ref{fig:randt}(c):
\begin{align}
  S^{\uparrow} &= S_{\rm f}^{+} - S_{\rm b}^{-} = t^{+} - r^{-} - 1, \label{eq:msup}\\
  S^{\downarrow} &= S_{\rm b}^{+} - S_{\rm f}^{-} = -t^{-} + r^{+} + 1. \label{eq:msdown}
\end{align}
A symmetric component of Eqs.~\eqref{eq:msup} and \eqref{eq:msdown}  
gives electric dipole radiation,
whereas an anti-symmetric component 
gives magnetic dipole radiation.
Therefore, we acquire
\begin{align}
\hat{\alpha}^{\rm em} &\propto S^{\uparrow} + S^{\downarrow} = t^{+} - t^{-} + r^{+} - r^{-}, \label{eq:aem}\\
\hat{\alpha}^{\rm mm} &\propto S^{\uparrow} - S^{\downarrow} = t^{+} + t^{-} - r^{+} - r^{-} -2. \label{eq:amm}
\end{align}
Equation \eqref{eq:aem} corresponds to Eq.~\eqref{eq:XYem} 
whereas Eq.~\eqref{eq:amm} correspond to Eq.~\eqref{eq:YYmm}.

\subsection{Classification of bianisotropic metasurfaces}
Reciprocity in the bianisotropic metasurfaces requires 
the following relations \cite{simovski_book}:
\begin{equation}
\hat{\alpha}_{xx}^{\rm em} = - \hat{\alpha}_{xx}^{\rm me}, \quad
\hat{\alpha}_{yy}^{\rm em} = - \hat{\alpha}_{yy}^{\rm me}, \quad
\hat{\alpha}_{xy}^{\rm em} = - \hat{\alpha}_{yx}^{\rm me}, \quad
\hat{\alpha}_{yx}^{\rm em} = - \hat{\alpha}_{xy}^{\rm me}. 
\label{eq:recipro2}
\end{equation}
Uniaxial metasurfaces with magneto-electric interactions
can be classified into four categories in terms of 
non-reciprocity and polarization rotation 
as shown in Table \ref{table:bamtsf}\cite{asadchy2018}.
In the following, 
we briefly explain each category.

\begin{table}[ht!]
\caption{Classification of bianisotropic metasurfaces}
\label{table:bamtsf}
\centering
\begin{tabular}{p{45mm} || p{35mm} | p{35mm}}
 & Reciprocal & Non-reciprocal \\
\hline \hline
With polarization  rotation 
& Chiral with $\tilde{t}_{yx}^{+} \neq \tilde{t}_{yx}^{-}$ 
& Tellegen with $\tilde{r}_{yx}^{+} \neq \tilde{r}_{yx}^{-}$ \\
\hline
Without polarization  rotation 
& Omega with $\tilde{r}_{xx}^{+} \neq \tilde{r}_{xx}^{-}$  
& Moving with $\tilde{t}_{xx}^{+} \neq \tilde{t}_{xx}^{-}$ \\ 
\end{tabular}
\end{table}

A reciprocal metasurface 
with polarization rotation 
is referred to as a chiral metasurface. 
In the chiral metasurface, 
the uniaxial structure and the reciprocity require 
transmission coefficients $\tilde{t}_{yx}^{\pm}$ to be 
\begin{equation}
\tilde{t}_{yx}^{+} - \tilde{t}_{yx}^{-} 
= -j \omega (\hat{\alpha}_{yy}^{\rm em} - \hat{\alpha}_{xx}^{\rm me})
= -j \omega (\hat{\alpha}_{yy}^{\rm em} - \hat{\alpha}_{yy}^{\rm me}) \neq 0 .
\label{eq:chiral}
\end{equation}
Equation \eqref{eq:chiral} indicates the optical activity, 
in which the transmission coefficients depend on propagation direction, 
and the polarization rotation is reversed for opposite propagating waves. 
Note that magnetized media or media under magnetic fields 
could show similar polarization rotation in transmission, 
known as the MO Faraday effects, 
with $\tilde{t}_{yx}^{+} \propto \hat{\alpha}_{yx}^{\rm ee}$ 
or $\tilde{t}_{yx}^{+} \propto \hat{\alpha}_{yx}^{\rm mm}$. 
However, 
these non-reciprocal MO effects 
do not show propagation direction dependence
with respect to polarization rotation direction.

The reflection coefficients with polarization rotation, 
$\tilde{r}_{yx}^{\pm}$, 
lead to
\begin{equation}
\tilde{r}_{yx}^{+} - \tilde{r}_{yx}^{-} 
= -j \omega (\hat{\alpha}_{yy}^{\rm em} + \hat{\alpha}_{xx}^{\rm me})
= -j \omega (\hat{\alpha}_{yy}^{\rm em} + \hat{\alpha}_{yy}^{\rm me}). 
\label{eq:tellegen}
\end{equation}
Equation \eqref{eq:tellegen}
becomes zero for the reciprocal chiral metasurfaces 
due to $\hat{\alpha}_{yy}^{\rm em} = - \hat{\alpha}_{yy}^{\rm me}$ 
as in Eq. \eqref{eq:recipro2},
but could be non-zero for non-reciprocal ones. 
When we compare reflection coefficients 
between $x$-polarized and $y$-polarized waves propagating in $+z$ direction
without mangeto-optical effects,
we obtain 
$\tilde{r}_{yx}^{+} 
= -j \omega (\hat{\alpha}_{yy}^{\rm em} + \hat{\alpha}_{xx}^{\rm me}) 
= - \tilde{r}_{xy}^{+}$.
The non-reciprocal metasurface with polarization rotation in reflection 
is referred to as a Tellegen metasurface, 
named after Bernard Tellegen \cite{tellegen1948}.
In reflection, 
the Tellegen metasurface shows 
polarization rotation in the same direction 
independent of polarization directions \cite{asadchy2018}. 
This is a non-reciprocal phenomenon 
with the broken time-reversal symmetry; 
therefore, the Tellegen metasurface seems to 
be similar to 
MO media mentioned in the above.
However, 
the Tellegen metasurface 
does not show any polarization rotations in the transmission.

The reflection coefficients without polarization rotation 
have the following relation,
\begin{equation}
\tilde{r}_{xx}^{+} - \tilde{r}_{xx}^{-} 
= -j \omega (\hat{\alpha}_{xy}^{\rm em} - \hat{\alpha}_{yx}^{\rm me}) .
\label{eq:omega}
\end{equation}
Equation \eqref{eq:omega}
becomes non-zero even in reciprocal metasurfaces
due to $\hat{\alpha}_{xy}^{\rm em} = - \hat{\alpha}_{yx}^{\rm me}$ 
as in Eq. \eqref{eq:recipro2}.
In this case,
reflection coefficients without polarization rotation depend
on incident direction \cite{niemi2013},
and the metasurface with  $\tilde{r}_{xx}^{+} \neq \tilde{r}_{xx}^{-}$ is referred to as 
an omega metasurface,
which is typically observed with omega-shaped metallic structures, 
such as SRRs.

The transmission coefficients without polarization rotation 
have the following relation,
\begin{equation}
  \tilde{t}_{xx}^{+} - \tilde{t}_{xx}^{-} 
  = -j \omega (\hat{\alpha}_{xy}^{\rm em} + \hat{\alpha}_{yx}^{\rm me}),
  \label{eq:moving}
  \end{equation}
which is always zero for reciprocal metasurfaces 
represented by 
$\hat{\alpha}_{xy}^{\rm em} = - \hat{\alpha}_{yx}^{\rm me}$ 
as in Eq. \eqref{eq:recipro2},
and could be non-zero for non-reciprocal metasurfaces.
The symmetric off-diagonal parts in magneto-electric interactions, 
i.e., $\hat{\alpha}_{xy}^{\rm em} = \hat{\alpha}_{yx}^{\rm me} \neq 0$
causes directional-dependent transmission coefficients, 
which correspond to the MCh effect.
The metasurface 
with $\tilde{t}_{xx}^{+} \neq \tilde{t}_{xx}^{-}$ 
is called a moving metasurface,
which exhibits an electromagnetic response 
equivalent to that of a regular material moving at a relativistic speed.

\section{Numerical simulation setup}
\begin{figure}[ht!]
\centering\includegraphics[width=13cm]{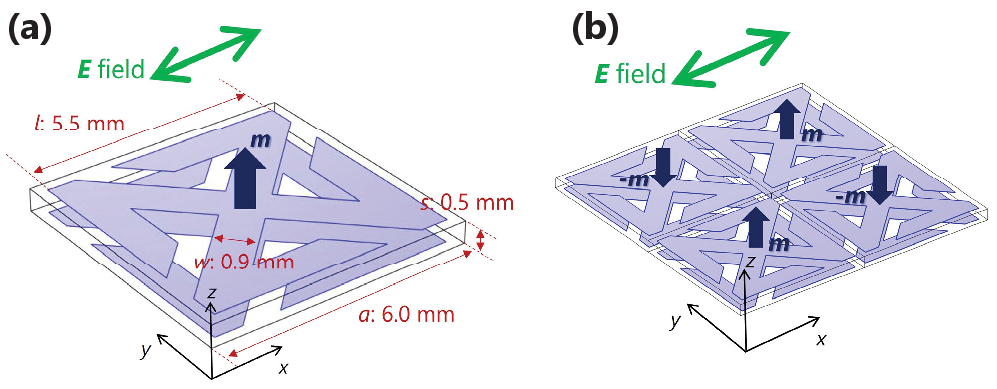}
\caption{
(a) A unit cell of a calculated MCh metasurface 
consisting of 5.5 mm $\times$ 5.5 mm Z-type gammadions 
on the both sides of a 0.5 mm-thick substrate.
The substrate is magnetized in the $+z$ direction.
The electric field of the incident microwave 
is polarized along the $x$ direction.
(b) A unit cell of a calculated composite metasurface 
including ordinary and contraposition MCh metasurfaces.}
\label{fig:mtsf}
\end{figure}

Figure \ref{fig:mtsf}(a) illustrates 
a unit cell of the uniaxial metasurface in this study. 
The metasurface consists of 
Z-type gammadions with $l =$ 5.5 mm  
on the both sides of a substrate with $s =$ 0.5 mm.
The design of the double Z-type gammadions are the same as 
those studied in Ref. \cite{kim2014}.
The gammadions are 
made of perfect electric conductor with $w =$ 0.9 mm, 
and in the unit cell with $a =$ 6.0 mm, 
which is repeated in the $x$ and $y$ directions by a periodic boundary condition.
The substrate is magnetized in the $+z$ direction.
The relative electric permittivity of the substrate 
is constant ($\varepsilon_{\rm r} = 3.3$).
Contrastingly, 
the relative magnetic permeability 
expressed by a tensor $\mu_{\rm r}$ 
is varied to express the magnetized substrates. 
The permeability tensor $\mu_{\rm r}$ for lossless substrates 
is described by 
\begin{equation}
\label{mu}
\mu_{\rm r} = 
\begin{bmatrix}
1 & j m & 0 \\
-j m & 1 & 0 \\
0 & 0 & 1
\end{bmatrix},
\end{equation}
where a parameter denoted by $m$ 
is related to magnetization of the substrates. 
The substrate is non-magnetic when $m = 0$.
On the other hand, 
$m > 0$ correspond to magnetic substrates magnetized in the $+z$ direction 
indicated as a navy arrow in Fig.~\ref{fig:mtsf}(a).
Non-zero $m$ causes 
non-zero off-diagonal components in $\mu_{\rm r}$,
resulting in MO effects. 

The $S$-parameters for microwave transmission and reflection
are numerically calculated using RF module of COMSOL Multiphysics.
The metasurface in the $x$-$y$ plane is placed 
at $z = 0$ in a vacuum ranging from $z = +$ 15 mm to $z = -$ 15 mm.
Ports 1 and 3 are set at $z = +$ 15 mm, 
while ports 2 and 4 are set at $z = -$ 15 mm.
The calculation space is terminated 
by 15 mm-long perfect matching layers 
at the both ends of the vacuum.
The microwaves 
in a frequency range between 4 and 16 GHz 
propagate in  $-z$ direction 
from port 1 to the metasurface.
Incident microwave 
is linearly polarized along the $x$-direction 
as indicated by a green arrow in Fig.~\ref{fig:mtsf}(a).
We numerically calculate 
$S_{\rm 11}$ detected by port 1 corresponding to reflection with $x$-polarization, 
$S_{\rm 21}$ detected by port 2 corresponding to transmission with $x$-polarization, 
$S_{\rm 31}$ detected by port 3 corresponding to reflection with $y$-polarization,
and $S_{\rm 41}$ detected by port 4 corresponding to transmission with $y$-polarization. 

The $S$-parameters $S_{\rm 11}$, $S_{\rm 21}$, $S_{\rm 31}$, and $S_{\rm 41}$ 
are converted to 
reflection and transmission coefficients of the metasurface
by using $r^-_{xx} = S_{\rm 11} e^{j k_0 d}$,
$t^-_{xx} = S_{\rm 21} e^{j k_0 d}$,
$r^-_{yx} = S_{\rm 31} e^{j k_0 d}$,
and $t^-_{yx} = S_{\rm 41} e^{j k_0 d}$,
where $k_0$ is the wavenumber in a vacuum 
and $d$ is the distance between port 1(3) and port 2(4).
In the same fashion, reflection and transmission coefficients 
for incident waves propagating in the $+z$ direction
can be calculated. 
By using these coefficients, 
we obtain each component of the effective polarizability tensor.

\section{Results and discussion}
\subsection{Transmission and reflection spectra of chiral and magnetochiral metasurfaces}
\begin{figure}[ht!]
\centering\includegraphics[width=13cm]{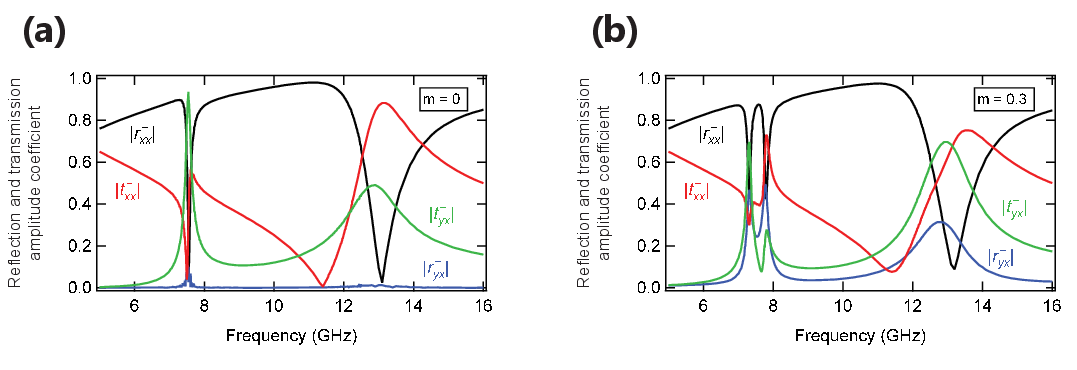}
\caption{
Calculated spectra of 
$|r_{xx}^{-}|$ (black), 
$|t_{xx}^{-}|$ (red), 
$|r_{yx}^{-}|$ (blue),
and $|t_{yx}^{-}|$ (green) 
in a frequency range of 4-16 GHz 
with (a) $m = 0$ and (b) $m = 0.3$.}
\label{fig:tr}
\end{figure}

Figure \ref{fig:tr}(a) 
presents calculated transmission and reflection amplitude spectra of 
$|r^-_{xx}|$ (black), 
$|t^-_{xx}|$ (red), 
$|r^-_{yx}|$ (blue), 
and $|t^-_{yx}|$ (green)
in a frequency region of 4-16 GHz 
for a chiral metasurface with $m = 0$.
$|r^-_{yx}|$ is almost zero 
in this frequency region. 
In the $|r^-_{xx}|$ spectrum, 
a sharp dip appears at 7.5 GHz, 
while a broad dip appears at 13.1 GHz.
At each frequency,
$|t^-_{xx}|$ spectrum 
exhibits dispersion-type features. 
These are caused by resonances 
by the double Z-type gammadions. 
The resonance in the gammadion structures 
results in polarization rotation of linearly polarized microwave,
i.e., optical activity.
Indeed, in Fig.~\ref{fig:tr}(a), 
the $|t^-_{yx}|$ spectrum presents 
a sharp peak at 7.5 GHz, 
and a broad peak at 13.1 GHz, 
highlighting the optical activity 
by the double Z-type gammadions \cite{kim2014}.

Figure \ref{fig:tr}(b) shows 
calculated spectra of
$|r^-_{xx}|$ (black), 
$|t^-_{xx}|$ (red), 
$|r^-_{yx}|$ (blue), 
and $|t^-_{yx}|$ (green)
for an MCh metasurface with $m = 0.3$.
In contrast to Fig.~\ref{fig:tr}(a), 
$|r_{yx}^{-}|$ in Fig.~\ref{fig:tr}(b) shows peaks at approximately 7.5 and 13.1 GHz.
In all spectra, 
the resonant mode around 7.5 GHz 
is split into two mods at 7.3 and 7.8 GHz.
By introducing 
perpendicular magnetization in the substrate, 
the MCh effect is induced. 
To verify the MCh effect, 
we calculate transmission phase spectra.

\begin{figure}[ht!]
\centering\includegraphics[width=7cm]{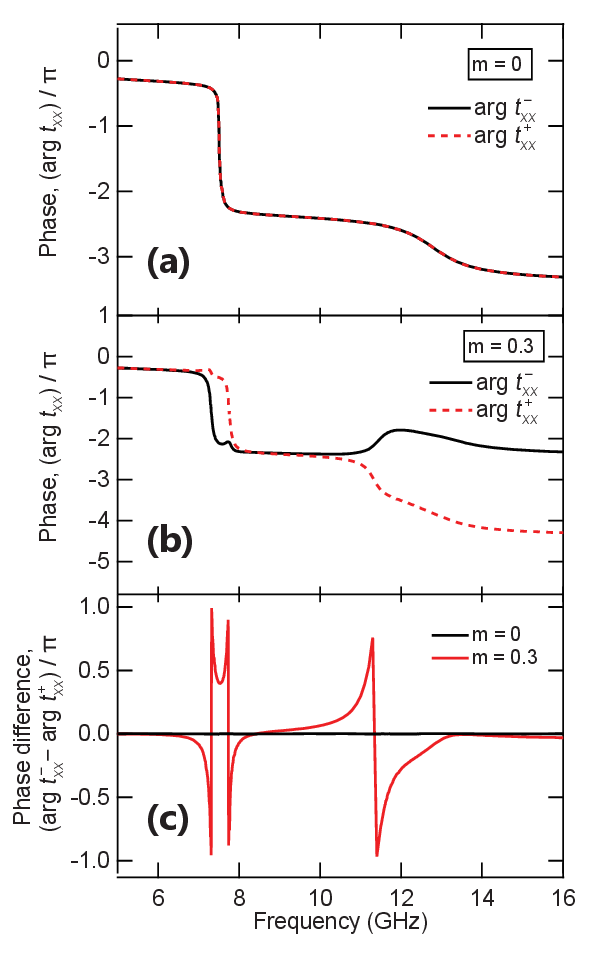}
\caption{
Calculated spectra of 
(arg $t_{xx}^{-}$)/$\pi$ (black solid) and (arg $t_{xx}^{+}$)/$\pi$ (red dashed) 
with (a) $m = 0$ and (b) $m = 0.3$
in a frequency range of 4-16 GHz .
(c) (arg $t_{xx}^{-}$ $-$ arg $t_{xx}^{+}$)/$\pi$ spectra 
with $m = 0$ (black) and $m = 0.3$ (red).}
\label{fig:phi}
\end{figure}

Figure \ref{fig:phi}(a) presents 
phase spectra of the chiral metasurface with $m = 0$.
The phase spectrum (arg $t_{xx}^{-}$)/$\pi$ (black solid) 
corresponds to the transmission in the $-z$ direction  
while (arg $t_{xx}^{+}$)/$\pi$ (red dashed) 
corresponds to the transmission in the +$z$ direction.
In Fig. \ref{fig:phi}(a), 
(arg $t_{xx}^{-}$)/$\pi$ and (arg $t_{xx}^{+}$)/$\pi$ are identical, 
although the phase shift is observed at 7.5 GHz and 13.1 GHz 
due to resonances in the gammadion structures.
However, 
by introducing magnetization with $m = 0.3$, 
a phase difference appears as in Fig. \ref{fig:phi}(b).
In Fig. \ref{fig:phi}(b), 
(arg $t_{xx}^{-}$)/$\pi$ (black solid)  
is different from (arg $t_{xx}^{+}$)/$\pi$ (red dashed) 
in regions of 7-8 GHz and 10-13 GHz.
Figure \ref{fig:phi}(c) represents 
directional phase difference spectra (arg $t_{xx}^{-}$ $-$ arg $t_{xx}^{+}$)/$\pi$  
with $m = 0$ (black) and $m = 0.3$ (red). 
While no phase difference is observed with $m = 0$, 
a large phase difference is observed with $m = 0.3$.
The phase difference indicates that 
the MCh effect is induced 
in the double Z-type gammadions with perpendicular magnetization.
 
\subsection{Effective polarizability tensor of chiral metasurfaces}
\begin{figure}[ht!]
\centering\includegraphics[width=13cm]{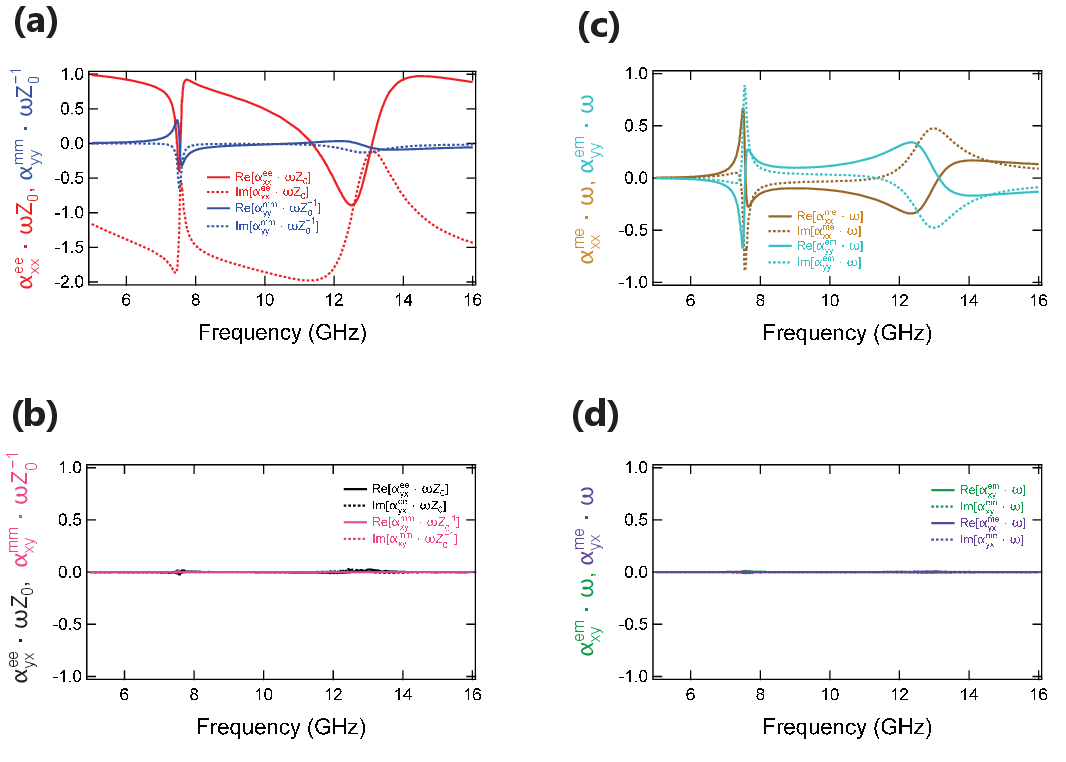}
\caption{Effective polarizabilities of chiral metasurfaces with $m = 0$ 
evaluated from numerical simulation. 
Frequency versus extracted 
(a) $\alpha_{xx}^{\rm ee} \omega Z_0$ (red) and $\alpha_{yy}^{\rm mm} \omega Z_0^{-1}$ (blue), 
(b) $\alpha_{yx}^{\rm ee} \omega Z_0$ (black) and $\alpha_{xy}^{\rm mm} \omega Z_0^{-1}$ (pink), 
(c) $\alpha_{xx}^{\rm me} \omega$ (gold) and $\alpha_{yy}^{\rm em} \omega$ (cyan),
(d) $\alpha_{xy}^{\rm em} \omega$ (green) and $\alpha_{yx}^{\rm me} \omega$ (purple).
Solid and dotted lines correspond to real and imaginary parts, respectively.}
\label{fig:chiral}
\end{figure}

The computed transmission and reflection coefficients 
are transformed to effective polarizabilities.
Figure \ref{fig:chiral} shows 
frequency spectra of 
the extracted effective polarizabilities
of the chiral metasurface with $m = 0$. 
Solid and dotted lines 
correspond to the real and imaginary parts, respectively.
In Fig.~\ref{fig:chiral}(a),
$\alpha_{xx}^{\rm ee} \omega Z_0$  (red) 
corresponds to the diagonal part of electric susceptibility 
while $\alpha_{yy}^{\rm mm} \omega Z_0^{-1}$ (blue) 
corresponds to the diagonal part of magnetic susceptibility.
The $\alpha_{xx}^{\rm ee} \omega Z_0$ spectrum 
representing electric dipole oscillation
exhibits a sharp resonance feature at 7.5 GHz 
and broad feature at 13 GHz.
The $\alpha_{yy}^{\rm mm} \omega Z_0^{-1}$ spectrum 
representing magnetic dipole oscillation
exhibits a sharp resonance at 7.5 GHz 
and very broad feature at 13 GHz.
Figure \ref{fig:chiral}(a) demonstrates that 
electric dipole oscillation is dominant at 13 GHz 
although magnetic dipole coexist.

Figure \ref{fig:chiral}(b) presents 
the off-diagonal part of electric susceptibility, $\alpha_{yx}^{\rm ee}  \omega Z_0$ (black), 
and of magnetic susceptibility, $\alpha_{xy}^{\rm mm} \omega Z_0^{-1}$ (pink).
The spectra of 
$\alpha_{yx}^{\rm ee}  \omega Z_0$ and 
$\alpha_{xy}^{\rm mm} \omega Z_0^{-1}$ 
indicate that 
the double Z-type gammadion with a non-magnetic substrate 
has no MO effects. 
In Fig.~\ref{fig:chiral}(c), 
$\alpha_{xx}^{\rm me} \omega$ (gold) 
and $\alpha_{yy}^{\rm em} \omega$ (cyan) correspond to 
chiral or Tellegen components.
The effective polarizabilities 
$\alpha_{xx}^{\rm me} \omega$ and $\alpha_{yy}^{\rm em} \omega$ 
show resonant features at 7.5 and 13 GHz.
We notice here $\alpha_{xx}^{\rm me} = - \alpha_{yy}^{\rm em}$.
Given $\alpha_{yy}^{\rm em} = \alpha_{xx}^{\rm em}$ 
due to the rotational symmetry, 
$\alpha_{xx}^{\rm me} = - \alpha_{xx}^{\rm em}$ is satisfied 
as Eq. \eqref{eq:recipro2};  
this is the hallmark of the reciprocal chiral-type bianisotropy.
This way, 
Fig.~\ref{fig:chiral}(c) demonstrates that 
the double Z-type gammadion
with the non-magnetic substrate 
shows polarization rotation 
by reciprocal chiral-type bianisotropy. 
This is consistent with the fact 
that $|r^-_{yx}| \simeq 0$ 
is observed in Fig.~\ref{fig:tr}(a).
In Fig.~\ref{fig:chiral}(d), 
$\alpha_{xy}^{\rm em} \omega$ (green) and $\alpha_{yx}^{\rm me} \omega$ (purple) 
correspond to omega or moving components.
Figure \ref{fig:chiral}(d) point out that 
moving and omega components are absent.

\subsection{Effective polarizability tensor of magnetochiral metasurfaces}
\begin{figure}[ht!]
\centering\includegraphics[width=13cm,clip]{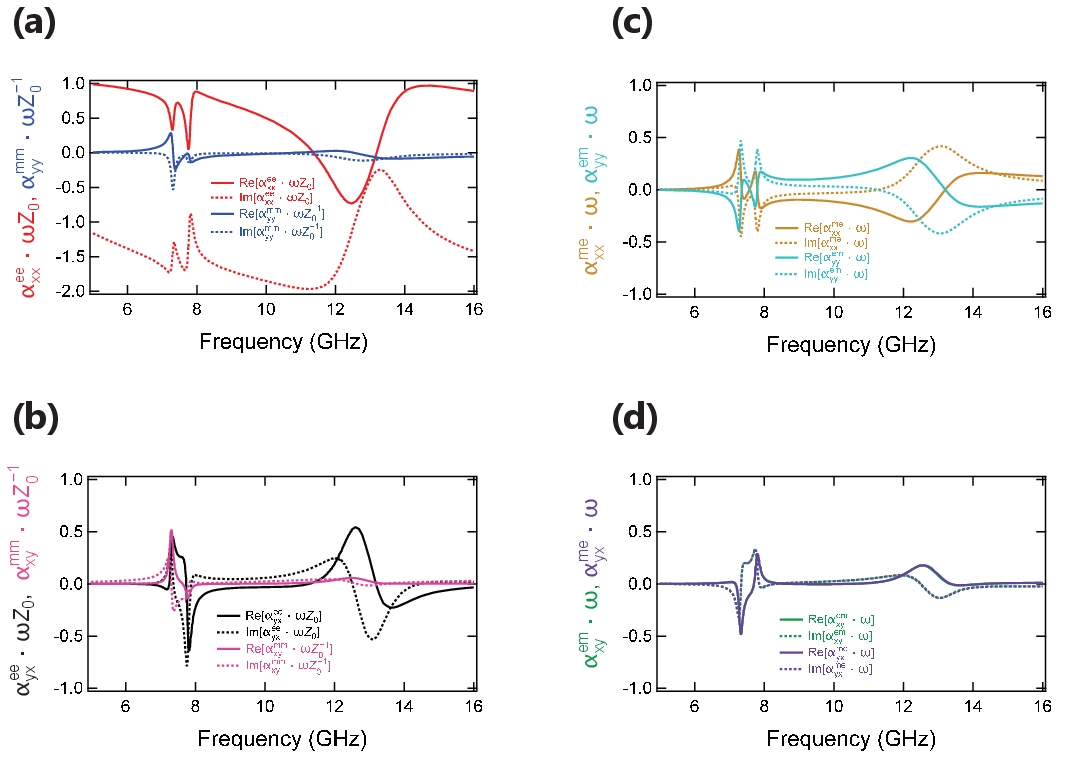}
\caption{Effective polarizabilities of magnetochiral metasurfaces with $m = 0.3$
evaluated from numerical simulation. 
Frequency versus extracted 
(a) $\alpha_{xx}^{\rm ee} \omega Z_0$ (red) and $\alpha_{yy}^{\rm mm} \omega Z_0^{-1}$ (blue), 
(b) $\alpha_{yx}^{\rm ee} \omega Z_0$ (black) and $\alpha_{xy}^{\rm mm} \omega Z_0^{-1}$ (pink), 
(c) $\alpha_{xx}^{\rm me} \omega$ (gold) and $\alpha_{yy}^{\rm em} \omega$ (cyan),
(d) $\alpha_{xy}^{\rm em} \omega$ (green) and $\alpha_{yx}^{\rm me} \omega$ (purple) .
Solid and dotted lines correspond to real and imaginary parts, respectively.}
\label{fig:mch}
\end{figure}

Figure \ref{fig:mch} shows 
extracted effective polarizabilities of the MCh metasurface with $m = 0.3$. 
Solid and dotted lines correspond to real and imaginary parts, respectively.
Figure \ref{fig:mch}(a) presents that
the resonant mode originally at 7.5 GHz with $m = 0$ 
is split into two modes at 7.3 and 7.8 GHz with $m = 0.3$.
Figure \ref{fig:mch}(c) 
presents that the double Z-type gammadion
with perpendicularly magnetized substrate 
exhibits reciprocal chiral-type bianisotropy, 
i.e., optical activity at 7.3, 7.8 and 13 GHz,
which is similar to Fig.~\ref{fig:chiral}(c). 
Contrastingly, 
by introducing magnetization in the substrate,
Figs.~\ref{fig:mch}(b) and \ref{fig:mch}(d) 
are much different from 
Fig.~\ref{fig:chiral}(b) and Fig.~\ref{fig:chiral}(d). 

In Fig.~\ref{fig:mch}(b), 
the off-diagonal part of electric susceptibility $\alpha_{yx}^{\rm ee} \omega Z_0$ (black) and 
that of magnetic susceptibility $\alpha_{xy}^{\rm mm} \omega Z_0^{-1}$ (pink) 
present resonant features at approximately 7.5 and 13 GHz, 
indicating the MO effects 
due to the time-reversal symmetry breaking.
Moreover,
Fig.~\ref{fig:mch}(d) shows 
non-zero $\alpha_{xy}^{\rm em} \omega$ (green) 
and $\alpha_{yx}^{\rm me} \omega$ (purple), 
which correspond to moving or omega-type bianisotropy.
Equation \eqref{eq:recipro2}, i.e., 
$\alpha_{xy}^{\rm em} \omega = \alpha_{yx}^{\rm me} \omega$ 
is satisfied in Fig.~\ref{fig:mch}(d); 
therefore,
the resonant features at 7.5 and 13 GHz 
are originated from 
the non-reciprocal moving-type bianisotropy.
Effective polarizability tensor analysis 
shown in Fig.~\ref{fig:mch} 
reveals that 
the double Z-type gammadion
with the perpendicularly magnetized substrate 
exhibits non-reciprocal moving-type bianisotropy 
without polarization rotation 
as well as optical activity and MO effect 
involving polarization rotation.
The non-reciprocal moving-type bianisotropy 
observed in Fig.~\ref{fig:mch}(d)
includes cascaded MCh effect \cite{rikken1998},
which is a combination of the optical activity and MO effect.

\subsection{Pure moving optical media with composite magnetochiral metasurfaces}
To obtain pure moving metasurfaces,
we combine the ordinary MCh metasurfaces studied in the above 
with contraposition MCh metasurfaces 
having both double inverse-Z-type gammadions 
and oppositely magnetized substrates.
Figure \ref{fig:mtsf}(b) illustrates 
a unit cell of the composite metasurfaces 
consisting of ordinary and contraposition MCh metasurfaces. 
The optical activity is cancelled 
by introducing contraposition MCh metasurfaces 
with double inverse-Z-type gammadion having inverse chirality
into the composite, 
similar to racemate of chiral chemical compounds.
Additionally, 
as shown in Fig.~\ref{fig:mtsf}(b),  
the contraposition metasurfaces have inverse magnetization in the $-z$ direction 
while the ordinary metasurfaces have magnetization in the $+z$ direction.
Therefore, the MO effect by the $-z$ direction magnetization 
cancels out the MO effect by the $+z$ direction magnetization, 
similar to antiferromagnets.
Nevertheless, 
the MCh effect is an odd function of chirality and magnetization direction; 
therefore, 
the contraposition metasurfaces 
has the identical sign of the MCh effect 
with the ordinary metasurfaces.
This way, 
combination between ordinary and contraposition MCh metasurfaces 
is anticipated to exhibit a pure moving effect 
with neither the optical activity nor MO effect. 

\begin{figure}[ht!]
\centering\includegraphics[width=7cm]{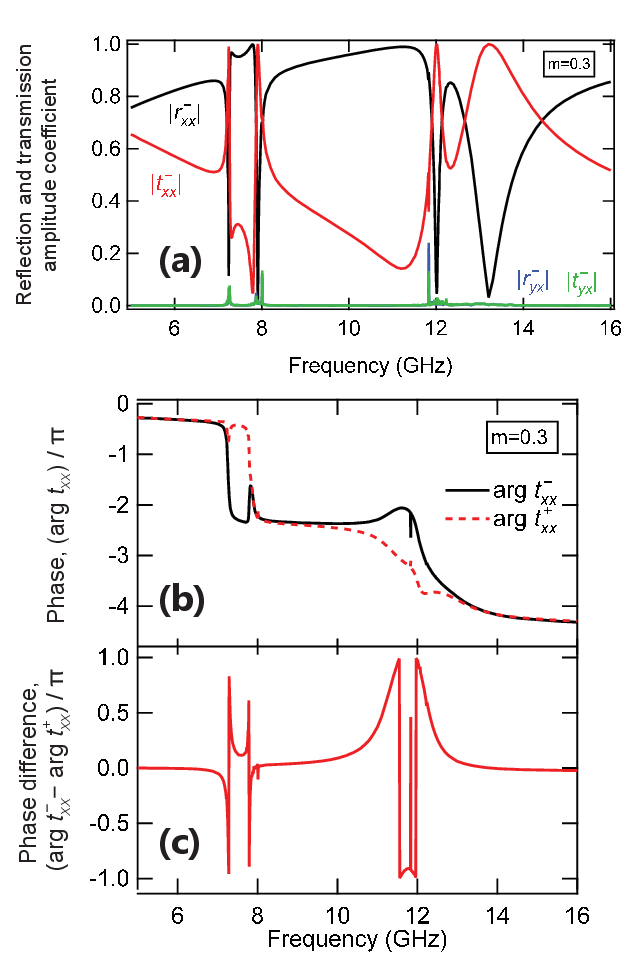}
\caption{(a) Calculated spectra 
in a frequency range of 4-16 GHz of 
$|r_{xx}^{-}|$ (black), 
$|t_{xx}^{-}|$ (red), 
$|r_{yx}^{-}|$ (blue),
and $|t_{yx}^{-}|$ (green) 
for a composite MCh metasurface 
consisting of the  ordinary MCh metasurfaces with $m = + 0.3$ 
and contraposition MCh metasurfaces with $m = - 0.3$. 
(b) Calculated spectra of 
(arg $t_{xx}^{-}$)/$\pi$ (black solid) 
and (arg $t_{xx}^{+}$)/$\pi$ (red dashed) , 
and (c)  (arg $t_{xx}^{-}$ - arg $t_{xx}^{+}$)/$\pi$
in a frequency range of 4-16 GHz 
for the composite MCh metasurface.}
\label{fig:contmch-rtphi}
\end{figure}

Figure \ref{fig:contmch-rtphi}(a) presents 
calculated spectra of 
$|r_{xx}^{-}|$ (black), 
$|t_{xx}^{-}|$ (red), 
$|r_{yx}^{-}|$ (blue),
and $|t_{yx}^{-}|$ (green) 
of the composite MCh metasurface with $m = \pm 0.3$ 
in a frequency range of 4-16 GHz. 
In Fig. \ref{fig:contmch-rtphi}(a),
$|r_{xx}^{-}|$ (black) and $|t_{xx}^{-}|$ (red)  
show resonant features 
at 7.2, 7.8, 12.0, and 13.2 GHz.
Moreover, 
$|r_{yx}^{-}|$ (blue) and $|t_{yx}^{-}|$ (green) 
accompanied by polarization rotation 
are almost zero, 
indicating that 
both the chiral-type bianisotropy and MO effect with polarization rotation 
are no longer functioning in the composite metasurface.
Contrastingly, 
the moving-type bianisotropy survives 
as shown in Figs. \ref{fig:contmch-rtphi}(b) and \ref{fig:contmch-rtphi}(c).
Figure \ref{fig:contmch-rtphi}(b) shows 
transmission phase spectra of 
(arg $t_{xx}^{-}$)/$\pi$ (black solid) 
and (arg $t_{xx}^{+}$)/$\pi$ (red dashed) 
of the composite metasurfaces.
Phase differences 
are observed at approximately 7.5 GHz and 12 GHz.
The directional phase difference spectrum,
(arg $t_{xx}^{-}$ $-$ arg $t_{xx}^{+}$)/$\pi$, 
plotted in Fig. \ref{fig:contmch-rtphi}(c) 
demonstrates the moving-type bianisotropy
at approximately 7.5 GHz and 12 GHz.

\begin{figure}[ht!]
\centering\includegraphics[width=13cm]{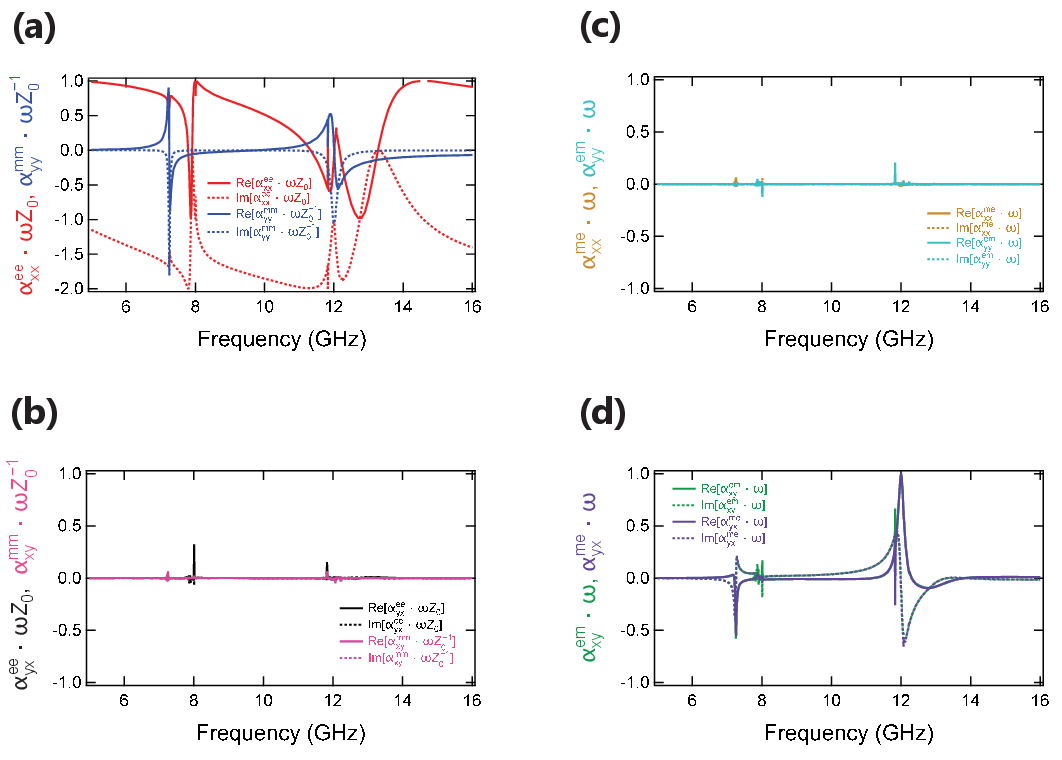}
\caption{Effective polarizabilities of composite MCh metasurfaces
evaluated from numerical simulation. 
Frequency versus extracted 
(a) $\alpha_{xx}^{\rm ee} \omega Z_0$ (red) and $\alpha_{yy}^{\rm mm} \omega Z_0^{-1}$ (blue), 
(b) $\alpha_{yx}^{\rm ee} \omega Z_0$ (black) and $\alpha_{xy}^{\rm mm} \omega Z_0^{-1}$ (pink), 
(c) $\alpha_{xx}^{\rm me} \omega$ (gold) and $\alpha_{yy}^{\rm em} \omega$ (cyan),
(d) $\alpha_{xy}^{\rm em} \omega$ (green) and $\alpha_{yx}^{\rm me} \omega$ (purple) 
of the composite metasurfaces 
consisting of ordinary MCh metasurfaces with $m = + 0.3$ 
and contraposition MCh metasurfaces with $m = - 0.3$.
Solid and dotted lines correspond to real and imaginary parts, respectively.}
\label{fig:contmch-spe}
\end{figure}

The survival of the moving-type bianisotropy 
is confirmed by effective polarizability tensor.
Figure \ref{fig:contmch-spe} shows 
extracted 
(a) $\alpha_{xx}^{\rm ee} \omega Z_0$ (red) and $\alpha_{yy}^{\rm mm} \omega Z_0^{-1}$ (blue), 
(b) $\alpha_{yx}^{\rm ee} \omega Z_0$ (black) and $\alpha_{xy}^{\rm mm} \omega Z_0^{-1}$ (pink), 
(c) $\alpha_{xx}^{\rm me} \omega$ (gold) and $\alpha_{yy}^{\rm em} \omega$ (cyan),
(d) $\alpha_{xy}^{\rm em} \omega$ (green) and $\alpha_{yx}^{\rm me} \omega$ (purple) 
of the composite MCh metasurface with $m = \pm 0.3$.
Magnetic dipole 
represented by $\alpha_{yy}^{\rm mm} \omega Z_0^{-1}$ 
in Fig.~\ref{fig:contmch-spe}(a)
shows a sharp resonant feature at 12 GHz
compared to the ordinary MCh metasurfaces 
shown in Figs. \ref{fig:mch}(a). 
The MO effect and chiral-type bianisotropy are eliminated 
as in Fig.~\ref{fig:contmch-spe}(b) and (c), respectively.
Contrastingly and most strikingly, 
Fig. \ref{fig:contmch-spe}(d) demonstrates 
the moving-type bianisotropy.
The optical moving effect at 12 GHz 
is much more enhanced than that at 7.5 GHz.

\begin{figure}[ht!]
\centering\includegraphics[width=7cm]{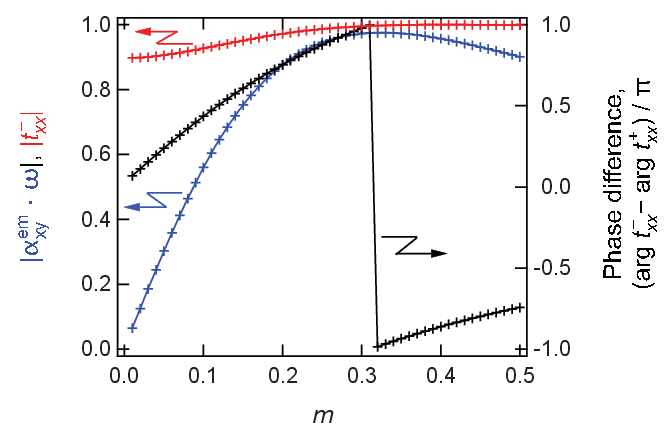}
\caption{$| \alpha_{xy}^{\rm em} \omega |$ (blue, left axis), 
$| t_{xx}^{-} |$ (red, left axis), and 
(arg $t_{xx}^{-}$ $-$ arg $t_{xx}^{+}$)/$\pi$ (black, right axis) at 12.0 GHz 
are plotted as a function of $m$ from 0 to 0.5.}
\label{fig:gyrator}
\end{figure}

The transmission and reflection amplitude spectra 
in Fig. \ref{fig:contmch-rtphi}(a) show that  
perfect transmission and no reflection are achieved at 12 GHz, 
indicating the impedance matching.
The impedance matching is expressed as
Re[$\alpha_{xx}^{\rm ee} \omega Z_0$] = Re[$\alpha_{yy}^{\rm mm} \omega Z_0^{-1}$] 
and Im[$\alpha_{xx}^{\rm ee} \omega Z_0$] = Im[$\alpha_{yy}^{\rm mm} \omega Z_0^{-1}$], 
corresponding to Kerker condition \cite{kerker1983},
which is satisfied at 12 GHz in Fig. \ref{fig:contmch-spe}(a).
We notice in Fig. \ref{fig:contmch-spe}(d) that
$| \alpha_{xy}^{\rm em} \omega |$ is maximized, 
i.e., $| \alpha_{xy}^{\rm em} \omega | = 1$, 
at 12 GHz.
The maximized $| \alpha_{xy}^{\rm em} \omega | = 1$ 
derives automatically 
both a perfect transparency and phase difference of $\pi$ 
as proved in the following.
In the moving medium, 
i.e., $\hat{\alpha}_{xy}^{\rm em} = \hat{\alpha}_{yx}^{\rm me}$,
Eq. \eqref{eq:moving} is written as
\begin{equation}
  \tilde{t}_{xx}^{+} - \tilde{t}_{xx}^{-} 
  = -2j \omega \hat{\alpha}_{xy}^{\rm em}.
  \label{eq:moving_max}
  \end{equation}
Absolute value of Eq. \eqref{eq:moving_max} is 
\begin{equation}
   2 |\omega \hat{\alpha}_{xy}^{\rm em}| 
   = |\tilde{t}_{xx}^{+} - \tilde{t}_{xx}^{-}|
   \leq |\tilde{t}_{xx}^{+}| + |\tilde{t}_{xx}^{-}|
   \leq 2.
  \label{eq:moving_max2}
  \end{equation}
Equation \eqref{eq:moving_max2} highlights that
the maximum value $| \alpha_{xy}^{\rm em} \omega | = 1$ is achieved 
when $\tilde{t}_{xx}^{+} = - \tilde{t}_{xx}^{-}$ and $|\tilde{t}_{xx}^{\pm}| = 1$.
The phase difference of $\pi$, 
(arg $t_{xx}^{-}$ $-$ arg $t_{xx}^{+}$) /$\pi = 1$, 
is indicated at 12 GHz in Fig. \ref{fig:contmch-rtphi}(c).

The effective polarizability tensor $| \alpha_{xy}^{\rm em} \omega |$,
transmission coefficient $| t_{xx}^{-} |$, and 
phase difference (arg $t_{xx}^{-}$ $-$ arg $t_{xx}^{+}$)/$\pi$ 
at 12.0 GHz are calculated and evaluated
with an increase in parameter $m$ 
related to magnetization from $m$ = 0 to 0.5.
In Fig. \ref{fig:gyrator},  
$| \alpha_{xy}^{\rm em} \omega |$ (blue, left axis),
$| t_{xx}^{-} |$ (red, left axis), and 
(arg $t_{xx}^{-}$ $-$ arg $t_{xx}^{+}$)/$\pi$ (black, right axis) 
at 12.0 GHz 
are plotted as a function of $m$.
Figure \ref{fig:gyrator} presents that 
$m$ = 0.31 brings about 
the maximum value of 
$| \alpha_{xy}^{\rm em} \omega | = 0.975$,
which is accompanied by 
(arg $t_{xx}^{-}$ $-$ arg $t_{xx}^{+}$)/$\pi$ = 1.0 
and $| t_{xx}^{-} | = 1.0$.
This way, 
the ideal gyrator for arbitrary polarizations is realized 
using the composite metasurface with ordinary and contraposition MCh metasurfaces.
Applications of the ideal gyrator include 
an isolator, circulator, 
non-reciprocal holography \cite{yang2023}, 
two-wavelength lasing \cite{vallet2001}, 
and artificial gauge field for light \cite{sawada2005}.
The pure moving optical medium 
is also relevant in fundamental physics 
in an analogy with a hypothetical particle called dyon \cite{khomskii2014}, 
and from the view point of interplay between physics and mathematics 
in algebraic geometry with Fresnel-Kummer surfaces \cite{hudson,favaro2016}.

\section{Conclusion}
Effective polarizability tensor is numerically evaluated 
to study bianisotropies in MCh metasurfaces.
The ordinary MCh metasurface 
consisting of the double Z-type gammadions 
with perpendicularly magnetized substrate 
shows the non-reciprocal moving-type bianisotropy 
as well as the reciprocal chiral-type bianisotropy 
and non-reciprocal MO effect. 
The chiral-type bianisotropy 
and MO effect 
are eliminated by the combining 
the ordinary MCh metasurfaces 
with the contraposition MCh metasurfaces 
having inverse-Z-type gammadion and oppositely magnetized substrates.
This realizes pure moving optical media.
The pure moving optical media 
demonstrate perfect transmission with phase difference of $\pi$,
which is a key to embody an ideal gyrator for arbitrary spatial and polarization modes.

\begin{backmatter}
\bmsection{Funding}
This work is financially supported by KAKENHI (23K13621) and JST-CREST (JPMJCR2102).

\bmsection{Acknowledgments}
The authors are grateful to Y. Tamayama and T. Ueda for fruitful discussions and comments.
H. Kurosawa is also acknowledged for instructing the numerical calculation prior to this work. 

\bmsection{Disclosures}
The authors declare no conflicts of interest.

\bmsection{Data availability} Data underlying the results presented in this paper are not publicly available at this time but may be obtained from the authors upon reasonable request.
\end{backmatter}



\begin{thebibliography}{99}
\newcommand{\enquote}[1]{``#1''}
\bibitem{LL} L. D. Landau, E. M. Lifshitz, and L. P. Pitaevski, {\it Electrodynamics of Continuous Media (2nd edition)} (Butterworth-Heinemann, Oxford, 1984).
\bibitem{Kong} J. A. Kong, {\it Electromagnetic Wave Theory} (EMW Publishing, Cambridge, 2005).
\bibitem{simovski_book} C. Simovski and S. Tretyakov, {\it An introduction to Metamaterials and Nanophotonics} (Cambridge University Press, 2020).
\bibitem{asadchy2018} V. S. Asadchy, A. D\'{i}az-Rubio, and S. A. Tretyakov, \enquote{Bianisotropic Metasurfaces: Physics and Applications,} {\protect\JournalTitle{Nanophotonics}} \textbf{7}, 1069 (2018). 
\bibitem{toyoda2015} S. Toyoda, N. Abe, S. Kimura, Y. H. Matsuda, T. Nomura, A. Ikeda, S. Takeyama, and T. Arima, \enquote{One-Way Transparency of Light in Multiferroic CuB$_2$O$_4$,} {\protect\JournalTitle{Phys. Rev. Lett.}} \textbf{115}, 267207 (2015).
\bibitem{kuzmenko2015} A. M. Kuzmenko, V. Dziom, A. Shuvaev, Anna Pimenov, M. Schiebl, A. A. Mukhin, V. Yu. Ivanov, I. A. Gudim, L. N. Bezmaternykh, and A. Pimenov, \enquote{Large directional optical anisotropy in multiferroic ferroborate,} {\protect\JournalTitle{Phys. Rev. B}} \textbf{92}, 184409 (2015).
\bibitem{fiebig2016review} M. Fiebig, T. Lottermoser, D. Meier and M. Trassin, \enquote{The evolution of multiferroics,} {\protect\JournalTitle{Nature Reviews Materials}} \textbf{1}, 16046 (2016). 
\bibitem{chen2005} X. Chen, B.-I. Wu, J. A. Kong, and T. M. Grzegorczyk, \enquote{Retrieval of the effective constitutive parameters of bianisotropic metamaterials,} {\protect\JournalTitle{Phys. Rev. E}} \textbf{71}, 046610 (2005).
\bibitem{kida2005} N. Kida, T. Yamada, M. Konoto, Y. Okimoto, T. Arima, K. Koike, H. Akoh, and Y. Tokura, \enquote{Optical Magnetoelectric Effect in a Submicron Patterned Magnet,} {\protect\JournalTitle{Phys. Rev. Lett.}} \textbf{94}, 077205 (2005).
\bibitem{plum2009} E. Plum, V. A. Fedotov, and N. I. Zheludev, \enquote{Extrinsic electromagnetic chirality in metamaterials,} {\protect\JournalTitle{J. Opt. A: Pure Appl. Opt.}} \textbf{11}, 074009 (2009).
\bibitem{li2009} Z. Li, K. Aydin, and E. Ozbay, \enquote{Determination of the effective constitutive parameters of bianisotropic metamaterials from reflection and transmission coefficients,} {\protect\JournalTitle{Phys. Rev. E}} \textbf{79}, 026610 (2009).
\bibitem{niemi2013} T. Niemi, A. O. Karilainen, and S. A. Tretyakov, \enquote{Synthesis of Polarization Transformers,} {\protect\JournalTitle{IEEE Trans. Antennas Propag.}} \textbf{61}, 3102 (2013).
\bibitem{pfeiffer2014prap} C. Pfeiffer, and A. Grbic, \enquote{Bianisotropic Metasurfaces for Optimal Polarization Control: Analysis and Synthesis,} {\protect\JournalTitle{Phys. Rev. Appl.}} \textbf{2}, 044011 (2014).
\bibitem{pfeiffer2014prl} C. Pfeiffer, C. Zhang, V. Ray, L. Jay Guo, and A. Grbic, \enquote{High Performance Bianisotropic Metasurfaces: Asymmetric Transmission of Light,} {\protect\JournalTitle{Phys. Rev. Lett.}} \textbf{113}, 023902 (2014).
\bibitem{odit2016} M. Odit, P. Kapitanova, P. Belov, R. Alaee, C. Rockstuhl, and Y. S. Kivshar, \enquote{Experimental realisation of all-dielectric bianisotropic metasurfaces,} {\protect\JournalTitle{Appl. Phys. Lett.}} \textbf{108}, 221903 (2016).
\bibitem{lavigne2021} G. Lavigne and C. Caloz, \enquote{Generalized Brewster effect using bianisotropic metasurfaces,} {\protect\JournalTitle{Opt. Express}} \textbf{29}, 11361 (2021).
\bibitem{kim2014} T. T. Kim, S. S. Oh, H.-S. Park, R. Zhao, S.-H. Kim, W. Choi, B. Min, and O. Hess, \enquote{Optical Activity Enhanced by Strong Inter-Molecular Coupling in Planar Chiral Metamaterials,} {\protect\JournalTitle{Scientific Reports}} \textbf{4}, 5864 (2014).
\bibitem{groenewege1962} M. P. Groenewege, \enquote{A theory of magneto-optical rotation in diamagnetic molecules of low symmetry,} {\protect\JournalTitle{Mol. Phys.}} \textbf{5}, 541 (1962).
\bibitem{portigal1971} D. L. Portigal and E. Burstein, \enquote{Magneto-spatial dispersion effects on the propagation of electro-magnetic radiation in crystals,} {\protect\JournalTitle{J. Phys. Chem. Solids}} \textbf{32}, 603 (1971).
\bibitem{baranova1977} N. B. Baranova, Yu. V. Bogdanov, and B. Ya. Zeldovich, \enquote{Electrical analog of the Faraday effect and other new optical effects in liquids,} {\protect\JournalTitle{Opt. Commun.}} \textbf{22}, 243 (1977).
\bibitem{markelov1977} V. A. Markelov, M. A. Novikov, and A. A. Turkin, \enquote{Experimental observation of a new nonreciprocal magneto-optical effect,} {\protect\JournalTitle{JETP Lett.}} \textbf{25}, 378 (1977) [{\protect\JournalTitle{Pis'ma Zh. Eksp. Teor. Fiz.}} \textbf{25}, 404 (1977)].
\bibitem{baranova1979} N. B. Baranova and B. Ya. Zeldovich, \enquote{Theory of a new linear magnetorefractive effect in liquids,} {\protect\JournalTitle{Mol. Phys.}} \textbf{38}, 1085 (1979).
\bibitem{wagni1982} G. Wagni\`{e}re and A. Meier, \enquote{The influence of a static magnetic field on the absorption coefficient of a chiral molecule,} {\protect\JournalTitle{Chem. Phys. Lett.}} \textbf{93}, 78 (1982).
\bibitem{wagni1984} G. Wagni\`{e}re, \enquote{Magnetochiral dichroism in emission. Photoselection and the polarization of transitions,} {\protect\JournalTitle{Chem. Phys. Lett.}} \textbf{110}, 546 (1984)
\bibitem{barron1984} L. D. Barron and J. Vrbancich, \enquote{Magneto-chiral birefringence and dichroism,} {\protect\JournalTitle{Mol. Phys.}} \textbf{51}, 715 (1984).
\bibitem{rikken1997} G. L. J. A. Rikken and E. Raupach, \enquote{Observation of magneto-chiral dichroism,} {\protect\JournalTitle{Nature}} \textbf{390}, 493 (1997).
\bibitem{kleindienst1998} P. Kleindienst and G. H. Wagni\`{e}re, \enquote{Interferometric detection of magnetochiral birefringence,} {\protect\JournalTitle{Chem. Phys. Lett.}} \textbf{288}, 89 (1998).
\bibitem{rikken1998} G. L. J. A. Rikken and E. Raupach, \enquote{Pure and cascaded magnetochiral anisotropy in optical absorption,} {\protect\JournalTitle{Phys. Rev. E}} \textbf{58}, 5081 (1998).
\bibitem{vallet2001} M. Vallet, R. Ghosh, A. Le Floch, T. Ruchon, F. Bretenaker, and J.-Y. Th\'{e}pot, \enquote{Observation of Magnetochiral Birefringence,} {\protect\JournalTitle{Phys. Rev. Lett.}} \textbf{87}, 183003 (2001).
\bibitem{train2008} C. Train, R. Gheorghe, V. Krstic, L.-M. Chamoreau, N. S. Ovanesyan, G. L. J. A. Rikken, M. Gruselle, and M. Verdaguer, \enquote{Strong magneto-chiral dichroism in enantiopure chiral ferromagnets,} {\protect\JournalTitle{Nature Mater.}} \textbf{7}, 729-734 (2008).
\bibitem{tomita2014} S. Tomita, K. Sawada, A. Porokhnyuk, and T. Ueda, \enquote{Direct Observation of Magnetochiral Effects through a Single Metamolecule in Microwave Regions,} {\protect\JournalTitle{Phys. Rev. Lett.}} \textbf{113}, 235501 (2014).
\bibitem{tomita2018} S. Tomita, H. Kurosawa, T. Ueda, and K. Sawada, \enquote{Metamaterials with magnetism and chirality',} {\protect\JournalTitle{J. Phys. D: Appl. Phys.}} \textbf{51}, 083001 (2018).
\bibitem{atzori2020} M. Atzori, G. L. J. A. Rikken, and C. Train, \enquote{Magneto-Chiral Dichroism: A Playground for Molecular Chemists,} {\protect\JournalTitle{Chem. Eur. J.}} \textbf{26}, 9784 (2020). 
\bibitem{caridad2021} J. M. Caridad, C. Tserkezis, J. E. Santos, P. Plochocka, M. Venkatesan, J. M. D. Coey, N, Asger Mortensen, G. L. J. A. Rikken, and V. Krsti\'{c}, \enquote{Detection of the Faraday Chiral Anisotropy,} {\protect\JournalTitle{Phys. Rev. Lett.}} \textbf{126}, 177401 (2021).
\bibitem{mirmoosa2014} M. S. Mirmoosa, Y. Ra'di, V. S. Asadchy, C. R. Simovski, and S. A. Tretyakov, \enquote{Polarizabilities of Nonreciprocal Bianisotropic Particles,} {\protect\JournalTitle{Phys. Rev. Appl.}} \textbf{1}, 034005 (2014).
\bibitem{alaee2015} R. Alaee, M. Albooyeh, M. Yazdi, N. Komjani, C. Simovski, F. Lederer, and C. Rockstuhl, \enquote{Magnetoelectric coupling in nonidentical plasmonic nanoparticles: Theory and applications,} {\protect\JournalTitle{Phys. Rev. B}} \textbf{91}, 115119 (2015).
\bibitem{yazdi2016} M. Yazdi and N. Komjani, \enquote{Polarizability Calculation of Arbitrary Individual Scatterers, Scatterers in Arrays, and Substrated Scatterers,} {\protect\JournalTitle{J. Opt. Soc. Am. B}} \textbf{33}, 491 (2016).
\bibitem{pozar} D. M. Pozar, {\it Microwave Engineering, 3rd. Ed.} (Wiley,  2005).
\bibitem{smith2006} D. R. Smith, J. Gollub, J. J. Mock, W. J. Padilla, and D. Schurig, \enquote{Calculation and measurement of bianisotropy in a split ring resonator metamaterial,} {\protect\JournalTitle{J. Appl. Phys.}} \textbf{100}, 024507 (2006).
\bibitem{kriegler2010} C. \'{E}. Kriegler, M. S. Rill, S. Linden, and M. Wegener, \enquote{Bianisotropic Photonic Metamaterials,} {\protect\JournalTitle{IEEE Journal of Selected Topics in Quantum Electronics}} \textbf{16}, 367 (2010). 
\bibitem{tellegen1948} B. D. Tellegen,  \enquote{The gyrator, a new electric network element,} {\protect\JournalTitle{Philips Res Rep}} \textbf{3}, 81-101 (1948). 
\bibitem{kerker1983} M. Kerker, D.-S. Wang, and C. Giles, \enquote{Electromagnetic scattering by magnetic spheres,} {\protect\JournalTitle{J. Opt. Soc. Am. A}} \textbf{73}, 765 (1983).
\bibitem{yang2023} W. Yang, J. Qin, J. Long,  W. Yan, Y. Yang, C. Li, E. Li, J. Hu, L. Deng, Q. Du, and L. Bi, \enquote{A self-biased non-reciprocal magnetic metasurface for bidirectional phase modulation,} {\protect\JournalTitle{Nat. Electron.}} \textbf{6}, 225-234 (2023).
\bibitem{sawada2005} K. Sawada and N. Nagaosa, \enquote{Optical Magnetoelectric Effect in Multiferroic Materials: Evidence for a Lorentz Force Acting on a Ray of Light,} {\protect\JournalTitle{Phys. Rev. Lett.}} \textbf{95}, 237402 (2005).
\bibitem{khomskii2014} D. I. Khomskii, \enquote{Magnetic monopoles and unusual dynamics of magnetoelectrics,} {\protect\JournalTitle{Nature Commun.}} \textbf{5}, 4793 (2014). 
\bibitem{hudson} R. W. H. T Hudson, {\it Kummer's Quartic Surface}, (the University Press, Cambridge, 1905).
\bibitem{favaro2016} A. Favaro and F. W. Hehl, \enquote{Light propagation in local and linear media: Fresnel-Kummer wave surfaces with 16 singular points,} {\protect\JournalTitle{Phys. Rev. A}} \textbf{93}, 013844 (2016).
\end{thebibliography}


\end{document}